\documentclass[12pt]{article}
\usepackage[utf8]{inputenc}
\usepackage[letterpaper, margin=1in]{geometry}
\usepackage{graphicx}

\usepackage{float}
\usepackage[hidelinks]{hyperref}
\usepackage [autostyle, english = american]{csquotes}
\MakeOuterQuote{"}
\graphicspath{ {./img/} }
\usepackage{amssymb}
\usepackage{subfigure}
\usepackage{upquote}
\usepackage{wrapfig}
\usepackage{color}
\usepackage{geometry}
\usepackage{algorithm}
\usepackage{algpseudocode}
\usepackage{amsmath}
\usepackage[numbers,sort&compress]{natbib}
\usepackage{graphicx}
\usepackage{authblk}

\title{Computational statistics of segregation and dislocation activities of hydrogen charged free surfaces and grain boundaries}
\author[1]{Matthew J. Melfi}
\author[1,2,*]{S. Mohadeseh Taheri-Mousavi}
\affil[1]{Department of Materials Science and Engineering, Carnegie Mellon University, 5000 Forbes Avenue, Pittsburgh, PA, 15213}
\affil[2]{Department of Mechanical Engineering, Carnegie Mellon University, 5000 Forbes Avenue, Pittsburgh, PA, 15213}
\affil[*]{Corresponding author: smtaherimousavi@cmu.edu}

\begin{document}

\maketitle

\begin{abstract}  
    Revealing statistics of H-defect interactions provides insights into significant ductility loss due to the particular strain partitioning in H-charged structural alloys. Experimental investigation of these interactions is extremely difficult, labor-intensive, and costly. Here, we used MD and GCMC simulations and studied H-diffusion deformation at polycrystalline scale with atomic resolution efficiently. To study H-free surface interactions, large pillars including all possible angles and planes of free surfaces were modeled. To study H-grain boundary interactions, several polycrystalline models containing comparable statistics of low and high angle grain boundaries were examined. We studied the statistics of H-segregation tendencies based on free surface angles and grain boundary types. Dislocation activities were also classified for these various types and total density and strength were compared and analyzed compared to H-free samples. In the free surface model, it was observed that H was evenly distributed along the model's surface. Although the dislocation density was reduced compared to H-free samples, localized bands of dislocations were produced. Additionally in the polycrystalline samples, it was concluded that H tends to segregate along grain boundaries with misorientation angles $\leq$$25^{\circ}$. However, specific misorientation angle interactions -namely $>$$45^{\circ}$- led to increased dislocation density in H-charged samples compared to their H-free counterparts. 
\end{abstract}

\section{Introduction}
\label{introduction}
Hydrogen embrittlement (HE) has been one of the most debated topics in high-strength metallic materials for more than a century\textsuperscript{\cite{JohnsonWilliam, Fukai1993}}. HE has led to a reduction in ductility, fracture resistance, and service life in high-strength metallic alloys. This phenomenon has severely affected several industries with H-rich environments, such as gas and oil, nautical, aerospace, aviation, and automotive fields. Consequently, research efforts to understand HE have substantially increased, with the number of publications on this topic growing exponentially over the past 25 years\textsuperscript{\cite{Zhou2022}}.

These studies proposed multiple theories to explain the HE phenomena.  Two main theories, H-enhanced decohesion (HEDE) and H-enhanced localized plasticity (HELP)\textsuperscript{\cite{Fukai1993, Zhou2022, Mai2021, Song2013, Dong2022, Yang2018, Zhao2022, Sun2021, Wasim2021, Luo2020, Koyama2020, Wan2019, Alvaro2019, Luo2018, Deng2018, Nygren2018, Xie2016, Robertson2015, Song2014, Matsumoto2009, Wen2004, Birnbaum2003, Lu2001, Katz2001, Delafosse2001, Lufrano1996, Birnbaum1994, Shih1988, Chen, Fukai2005, Moeini-Ardakani2021, Wu2023, Taketomi2010, Lee2016, Kim2022, Moreno-Gobbi2011, Wen2009, Matsui1979, Kirchheim2010, Chen2013, Barnoush2010, Robertson2001, Martin2019, Bond1987, Sofronis1995, Yao2023, Pfeil1926, Martin2012, Troiano2016, Li2022, Djukic2019}}, have shown the most amount of discussions. Generally, under deformation, cracks propagate either along grain or phase boundaries or inside grains, depending on which has a lower cohesive energy. In some H-charged samples, however, it was observed that cracks mainly propagate along the boundaries. According to HEDE, this phenomena is the consequence of a lowering of the cohesive energy enough to be lower than the bulk material. As the H concentration increases, the cohesive energy continues to decrease\textsuperscript{\cite{Martin2012, Troiano2016, Li2022, Djukic2019}}.

Recently, HELP has sparked significant discussions, yet a centralized definition of this theory and its underlying mechanisms remain elusive. Generally in H-free cracked surfaces, the samples have diffused plasticity around the crack surfaces, and under deformation, the crack tip becomes blunted. Interestingly, when the sample is charged with H, the plasticity distribution has been enhanced and changed to be a narrow band under the crack surfaces. Under deformation of the H-charged samples, this sharp crack remains sharp as it propagates through this narrow plasticity band. There are several proposed explanations aimed at shedding light on this theory. One of these is the concept of lowering the dislocation-dislocation repulsions due to H atmosphere surrounding dislocations. This altered stress field creates a shielding effect around the dislocations, modifying their interaction energy with various obstacles such as secondary phases, solute atoms, and other dislocations. This lowered interaction energy leads to increased dislocation mobility\textsuperscript{\cite{Birnbaum1994, Barnoush2010, Robertson2001, Martin2019, Bond1987, Sofronis1995}}. This proposed mechanism was debated in a computational study\textsuperscript{\cite{Song2014}}. Another proposed explanation involves the lowering of the dislocation formation energy. Consequently, this reduction enhances the nucleation rates of dislocation kink-pairs, regions where the dislocation line deviates from its straight path due to impurities, lattice defects, or other imperfections. If kink-pair formation is considered the rate-limiting step for dislocation motion, higher nucleation rates result in increased dislocation mobility\textsuperscript{\cite{Kirchheim2010, Chen2013, Kirchheim2007}}. A third explanation involves a reduction in stacking fault energy. A lower stacking fault energy leads to wider stacking faults, which can hinder the cross-slip of dislocations. This restriction forces slip to occur on a single plane, thereby increasing the likelihood of crack propagation\textsuperscript{\cite{Wu2023, Taketomi2010, Lee2016, Kim2022, Symons, Zhou20222, Zheng2020}}.

Numerous studies have utilized various experimental methods to gain insights into HE at different length scales. Tensile tests are commonly employed to compare the performance of H-free and H-charged samples. However, these tests do not provide detailed information about the specific location of H or H-defect interactions within the material\textsuperscript{\cite{Wan2019, Alvaro2019, Deng2018, Barnoush2010}}. Some researchers have utilized electron backscatter diffraction (EBSD) to understand defect interactions such as grain interactions\textsuperscript{\cite{NISHIKAWA2011, Laureys2016, Laureys2020, Moshtaghi2022, Fressengeas2018, Koyama2017, Kim2019}}. Although the source of plasticity, dislocation emissions, and interactions, cannot be tracked explicitly the total effect of dislocation density on deformation gradient can be captured by EBSD. Nevertheless, these efforts have yet to conclusively demonstrate plasticity below the crack surfaces or establish a clear connection between observed deformation mechanisms and microstructural features like grain boundary orientation or dislocation density, primarily due to their large scale. Similarly, the location of H remains elusive, validating its presence at each defect type cannot be tracked explicitly\textsuperscript{\cite{Luo2020, Luo2018, Raabe2018}}. Cryogenic Atom Probe Tomography (APT), the most advanced experimental method for investigating segregation tendencies, has been employed to study H-defect interactions. With this method, researchers can finally comprehend the precise location of H. However, because these experiments are conducted at cryogenic temperatures, H kinetics are frozen and the room temperature H-defects interactions are not yet thoroughly understood. Moreover, the APT samples are at angstrom scale and thus the high cost of this method does not allow statistical analysis of H-segregation tendencies. Furthermore, this method cannot be applied during deformation of H-charged samples to dynamically track H diffusion at various combinations of defects during deformation\textsuperscript{\cite{Zhao2022, Chen}}. As a result, the H-defect interactions at large scale remain elusive in experimental techniques. 

Given the uncertainties and challenges inherent in experimental methods in explicitly H tracking, numerical simulations such as molecular dynamics (MD), density functional theory (DFT), and Monte Carlo (MC) simulations have been applied to understand H-defect interactions. MD simulations have been used to study both equilibrium and non-equilibrium processes, while MC simulations are able to capture equilibrium state mechanisms. DFT, by contrast, offers an atomistic perspective on H-defect interactions at the quantum mechanical level\textsuperscript{\cite{Kim2018, Lu2023, Hajilou2020, Schleder2019, Zhu2023, Christmann1979, Christmann1974, Ferrin2012, Staykov2014}}. However, many simulation attempts have length scale limitations and mainly focused solely on a single defect type\textsuperscript{\cite{Dong2022, Matsumoto2009, Wen2004, Song2012, Lu2016}} or grain boundary\textsuperscript{\cite{Mai2021, Dong2022, Yang2018}}. For instance, single-grain investigations typically simulate only a few thousand atoms or a few nanometers in length. Similarly, simulations of grain boundary interactions often examine only a single grain boundary, despite typical samples containing multiple grain boundaries that may interact with one another\textsuperscript{\cite{Mai2021, Dong2022, Yang2018, Tehranchi2017}}. These simulation types also have their limitations. As mentioned, MD simulations can be performed for equilibrium and non-equilibrium processes but are limited in capturing long time-scale mechanisms such as large-scale diffusion. Conversely, while MC simulations can capture these diffusion mechanisms, they cannot account for non-equilibrium processes such as deformation. Furthermore, they are not efficient in capturing short range equilibrium such as slight relocation of atoms around H atoms when they are inserted during MC steps\textsuperscript{\cite{Moeini-Ardakani2021}}. While offering unparalleled accuracy in calculating the electronic and atomic structures of H-defect interactions, DFT is constrained by high computational costs, limiting its application to small systems of tens to hundreds of atoms. 

MC diffusion simulations at the atomistic scale have been conducted using various frameworks, which can be classified into two main categories: those that directly handle the kinetics of diffusion and those that do not. When the actual movements of atoms are not considered, the focus is on the outcome of diffusion, primarily captured by thermodynamics and semiclassical statistical physics. In cases where the kinetics of diffusion is addressed, the kinetic MC technique is the most viable approach. 

For simulations that do not handle atomic movements directly, different ensembles are considered based on the type of alloying elements and the target thermodynamic properties. For interstitial alloying elements, such as H, grand canonical Monte Carlo (GCMC) simulations are commonly used\textsuperscript{\cite{Moeini-Ardakani2021, Gai2016, Diarra2012}}. These simulations allow for the insertion or deletion of elements at a given chemical potential or partial pressure. It is noteworthy that in all these simulations, the indistinguishability of atoms is accounted for, while other quantum effects are ignored\textsuperscript{\cite{Moeini-Ardakani2021}}.

The combination of MD and MC techniques seems ideal for coupled diffusion-deformation problem such as the H one. However, in practice, the probabilistic nature of the MC scheme and the evolving number of degrees of freedom complicate the efficient implementation and parallelization of hybrid MC-MD techniques. Recently, there have been multiple attempts at creating efficient parallelized MC simulations. Sadigh et al.\textsuperscript{\cite{Sadigh2012}} designed and proposed a novel parallelization scheme capable of performing simultaneous MC moves based on domain decomposition. Yamakov\textsuperscript{\cite{Yamakov2016}} implemented parallel MD and semi-grand canonical MC simulations using this algorithm to efficiently model the behavior of substitutional alloys. Moeini-Ardakani et al.\textsuperscript{\cite{Moeini-Ardakani2021}} developed and implemented a new GCMC algorithm that is as scalable as MD while being capable of addressing non-pair and many-body potentials. With this library, simulations can be performed to reveal H-defect interactions that have previously eluded researchers.

By using the Moeini-Ardakani et al.\textsuperscript{\cite{Moeini-Ardakani2021}} library, two different models depicting various defect types will be generated to study H-defect interactions: free surfaces and grain boundaries. Each of these models will be used to reveal H-segregation tendencies and the H influence on underlying deformation mechanisms and mainly dislocation activities.

\section{Methods}
\subsection{Sample generation}
\hspace{\parindent}
A pillar consisting of a single grain was filled with Nickel (Ni) atoms in a (100) orientation, to replicate a free surface. This shape was deliberately chosen to encompass all possible angles and planes, rather than studying each one individually. Two different samples with the following dimensions were modeled: one sample with a diameter of 6 nm and a length of 30 nm, consisting of approximately 75,000 Ni atoms, while the other sample has a diameter of 20 nm and a length of 30 nm, consisting of approximately 830,000 Ni atoms. To study H grain boundary interactions, five different three-dimensional (3D) polycrystalline samples were created with twenty-seven randomly oriented grains. Each of these grain boundary samples has dimensions of 15 nm $\times$ 15 nm $\times$ 15 nm, with an average grain size of 5 nm, consisting of approximately 310,000 Ni atoms. The main distinction among these grain boundary samples lies in the varying grain orientations and thus grain boundary misorientation angles. By creating a single database across the five different samples, the resulting distribution provides a non-biased dataset of low misorientation angles ($\leq$15$^{\circ}$) and high misorientation angles ($>$15$^{\circ}$). In this dataset, 49.17\% of the grain boundaries are represented as low misorientation angles and 50.83\% as high misorientation angles. Periodic boundary conditions were applied to all directions for each grain boundary sample.

\subsection{Numerical Setup}
MD and GCMC simulations were performed using the library developed by Moeini-Ardakani et al.\textsuperscript{\cite{Moeini-Ardakani2021}} and LAMMPS code\textsuperscript{\cite{Plimpton1995}}. The embedded atom method (EAM) potential\textsuperscript{\cite{Angelo1995}} was adapted for Ni and Ni-H interactions. The Nose-Hoover thermostat was used to maintain an ambient temperature of 300 K during the NPT ensemble. The integration time steps were set to 1 fs for the H-free and 0.5 fs for the H-charged samples. Each sample was initially relaxed for 500 ps at this temperature under NPT ensemble. GCMC simulations on $\mu$VT ensemble was conducted, allowing for H atoms to be inserted and extracted from the samples. The chemical potential of H was set to -2.462 eV\textsuperscript{\cite{Koyama2020}}. 

For the deformation process, LAMMPS was utilized. Before deformation, each sample underwent an initial relaxation period of 200 ps within the NPT ensemble. In the free surface samples, a 100\% deformation was applied along the Z-axis. In the grain boundary samples, a 10\% deformation was applied along the X-axis. All deformations were executed at a constant strain rate of $10^9 s^{-1}$.

\subsection{Post-processing analaysis}
The H atom coordinates were converted into polar coordinates to evaluate H-segregation tendencies along the free surface samples, particularly across different rotational angles. The Z-coordinate was used to assess H-segregation along the length of the free surface samples. To investigate segregation tendencies along the grain boundaries, the common neighbor analysis modifier (CNA)\textsuperscript{\cite{Faken1994}} from the Open Visualization Tool (OVITO) was applied, using a cut-off distance of 1 nm\textsuperscript{\cite{Faken1994, Stukowski2010}}. This method identifies the atoms surrounding an H atom, enabling the determination of the surrounding grains and, consequently, the identification of the misorientation angle.

To analyze dislocation densities in the free surface samples, the Gaussian density method was employed through the construct surface modifier\textsuperscript{\cite{Krone2012}}. This was used to determine the pillar volume, while the dislocation analysis (DXA) modifier\textsuperscript{\cite{Stukowski2010-2}} in OVITO was applied to identify and analyze dislocations. 

For the grain boundary samples, polyhedral template matching (PTM)\textsuperscript{\cite{Larsen2016}} and grain segmentation\textsuperscript{\cite{Bonald2018}} modifiers were applied to identify grains within the polycrystalline model. DXA was then used to identify and analyze dislocations. The combined use of these modifiers effectively removes dislocations along the grain boundaries, isolating only those emitted within the grains.

In each of these models, the aforementioned modifiers were applied to each frame. However, when using DXA, the dislocation ID values change from frame to frame. To address this issue, a dislocation tracking system was implemented to track the development of dislocations across frames. This was achieved by comparing the dislocation core atoms between consecutive frames throughout the deformation process and calculating a Jaccard similarity score\textsuperscript{\cite{Tan2018, Madhikermi2016}}, as shown in Eq. (\ref{Jaccard}). In this context, $A$ and $B$ represent the lists of dislocation core atoms being compared, $f$ denotes the frame number, $i$ and $j$ refer to the dislocation IDs, and $m$ and $n$ indicate the final dislocation IDs in the corresponding frames.

\begin{equation} 
    J(A_{i,f}, B_{j,f+1}) = \frac{|A_{i,f} \cap B_{j,f+1}|}{|A_{i,f} \cup B_{j,f+1}|}, \quad 
    \begin{cases} 
        i = 0, 1, 2, \ldots, m 
        \\ j = 0, 1, 2, \ldots, n 
    \end{cases} 
    \label{Jaccard} 
\end{equation}

The maximum Jaccard similarity score was used to track dislocations from frame to frame. However, if no similarity existed between $A$ and $B$, the dislocation was classified as a new dislocation. In such cases, the head and tail -representing the two ends of the dislocation- were analyzed to determine their origins. This process involved selecting the closest dislocation core atom to both the head and tail, then expanding the selection to include all atoms within a 1 nm cut-off distance to identify the surrounding atoms. In the polycrystalline samples, these selected atoms were then used to identify the grains and misorientation angle. Additionally in the free surface samples, the atoms were converted to polar coordinates to evaluate emission tendencies across different rotational angles. These atoms were also analyzed to assess dislocation emission along the length of the free surface samples using the Z-coordinate.

\section{Results}
\subsection{Free Surface}
\label{calphad}
\begin{figure}[H]
    \centering
    \begin{subfigure}[]{}
        \centering
        \includegraphics[width=0.31\textwidth]{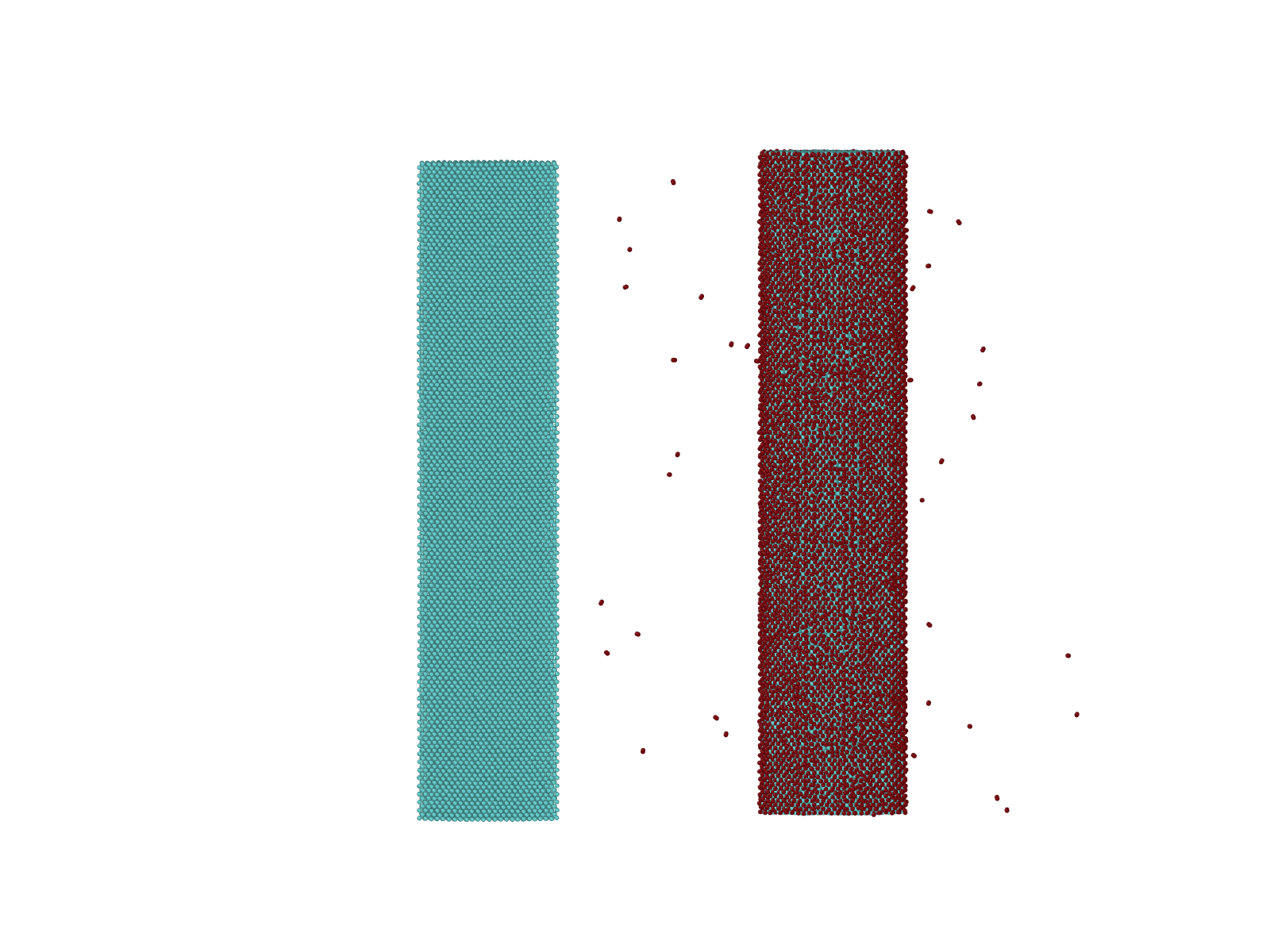}
    \end{subfigure}
    \begin{subfigure}[]{}
        \centering
        \includegraphics[width=0.31\textwidth]{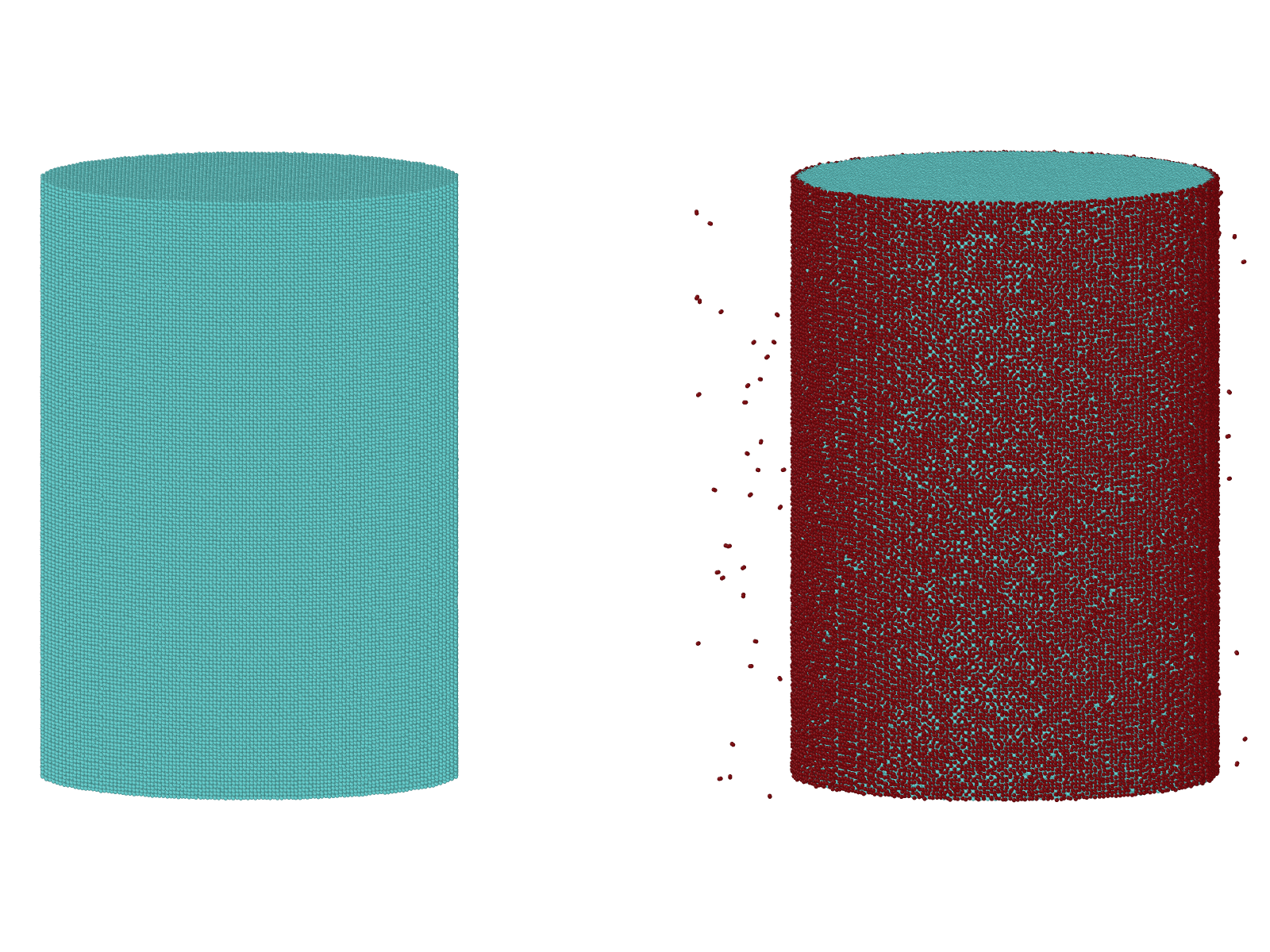}
    \end{subfigure}
    \begin{subfigure}[]{}
        \centering
        \includegraphics[width=0.31\textwidth]{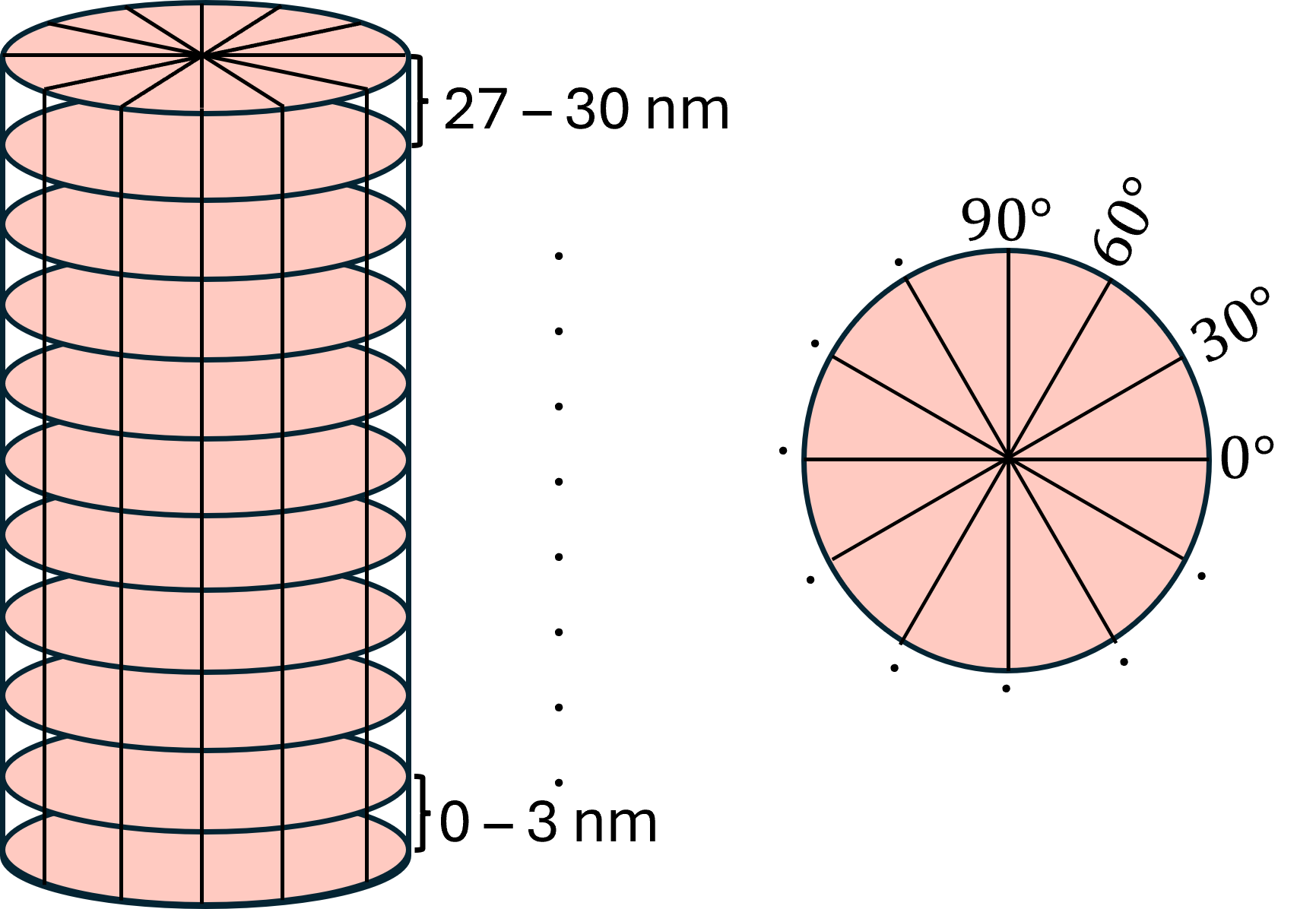}
    \end{subfigure}
   \caption{ H-free (left) and H-charged (right) samples with a 6 nm diameter and 30 nm length (a) and samples with a 20 nm diameter and 30 nm length (b). (c) Schematic illustrating various length and rotational angles.}
    \label{FS-samples}
\end{figure}

\begin{figure}[H]
    \centering
    \begin{subfigure}[]{}
        \centering
        \includegraphics[width=0.45\textwidth]{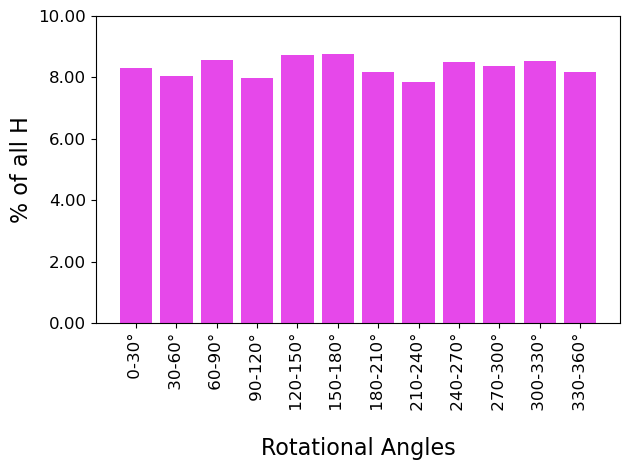}
        \label{FS-small-H-angle}
    \end{subfigure}
    \begin{subfigure}[]{}
        \centering
        \includegraphics[width=0.45\textwidth]{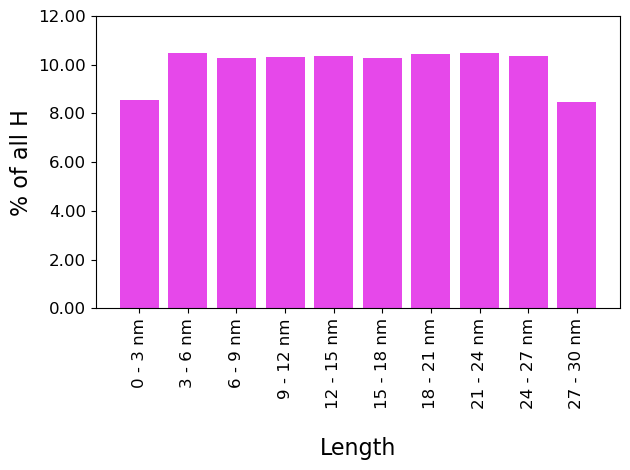}
        \label{FS-small-H-z}
    \end{subfigure}
    
    \begin{subfigure}[]{}
        \centering
        \includegraphics[width=0.45\textwidth]{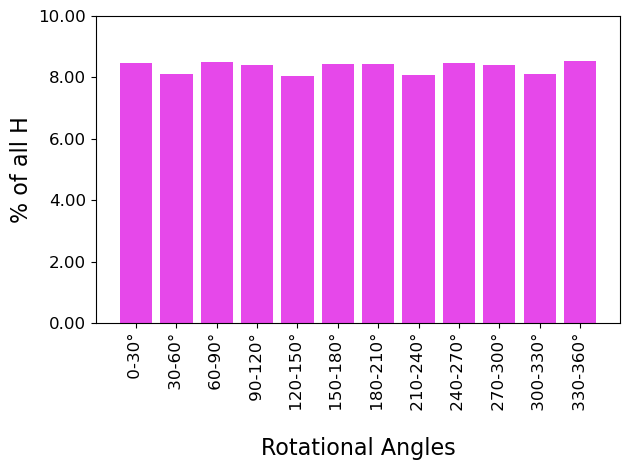}
        \label{FS-large-H-angle}
    \end{subfigure}
    \begin{subfigure}[]{}
        \centering
        \includegraphics[width=0.45\textwidth]{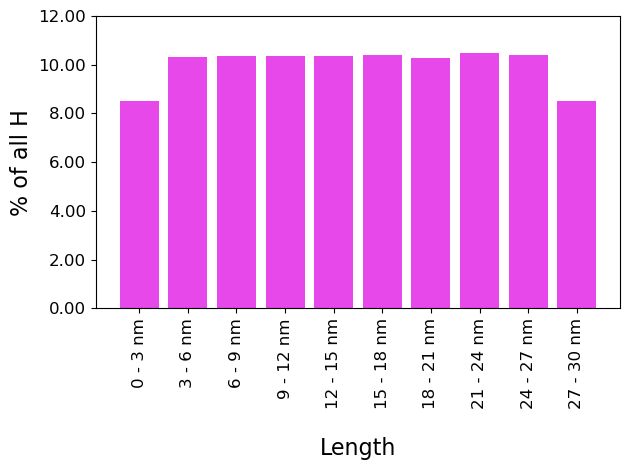}
        \label{FS-large-H-z}
    \end{subfigure}
    \caption{H-segregation along various rotational angles (a) and lengths (b) along the 6 nm diameter and 30 nm length sample and various rotational angles (c) and lengths (d) along the 20 nm diameter and 30 nm length sample.} 
    \label{FS-H}
    
\end{figure}

\hspace{\parindent}
As previously discussed, a pillar made of a single-grain crystal of Ni was used to simulate a free surface. Each sample underwent H-charging, as shown in Figure (\ref{FS-samples}a) and (\ref{FS-samples}b). This reveals that H tends to segregate along the free surface rather than penetrating deeper into the sample. To further explore this H-segregation, each sample was divided into rotational angles in 30$^{\circ}$ increments and subdivided into 3 nm sections along the Z-axis, Figure (\ref{FS-samples}c). As illustrated in Figure (\ref{FS-H}), H is uniformly distributed across the rotational angles and along the Z-axis and no particular angles and lengths had noticeable differences compared to others.

\begin{figure}[H]
    \centering
    \begin{subfigure}[]{}
        \centering
        \includegraphics[width=0.45\textwidth]{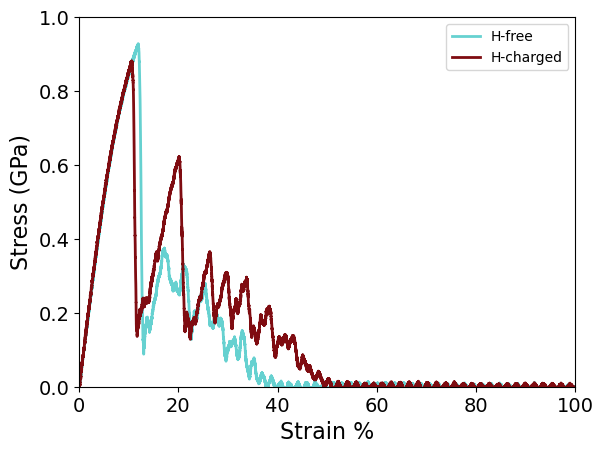}
    \end{subfigure}
    \begin{subfigure}[]{}
        \centering
        \includegraphics[width=0.45\textwidth]{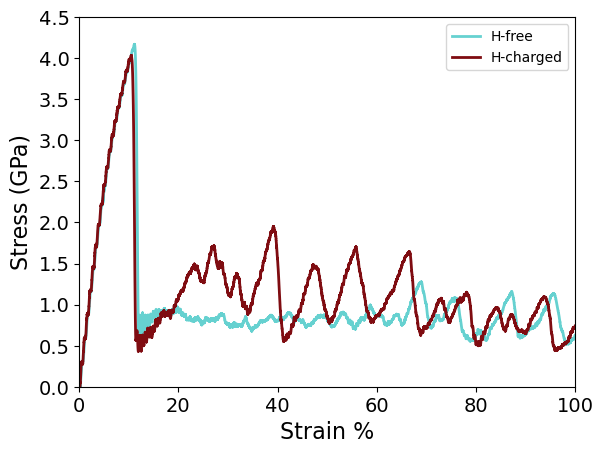}
    \end{subfigure}    
    
    \begin{subfigure}[]{}
        \centering
        \includegraphics[width=0.45\textwidth]{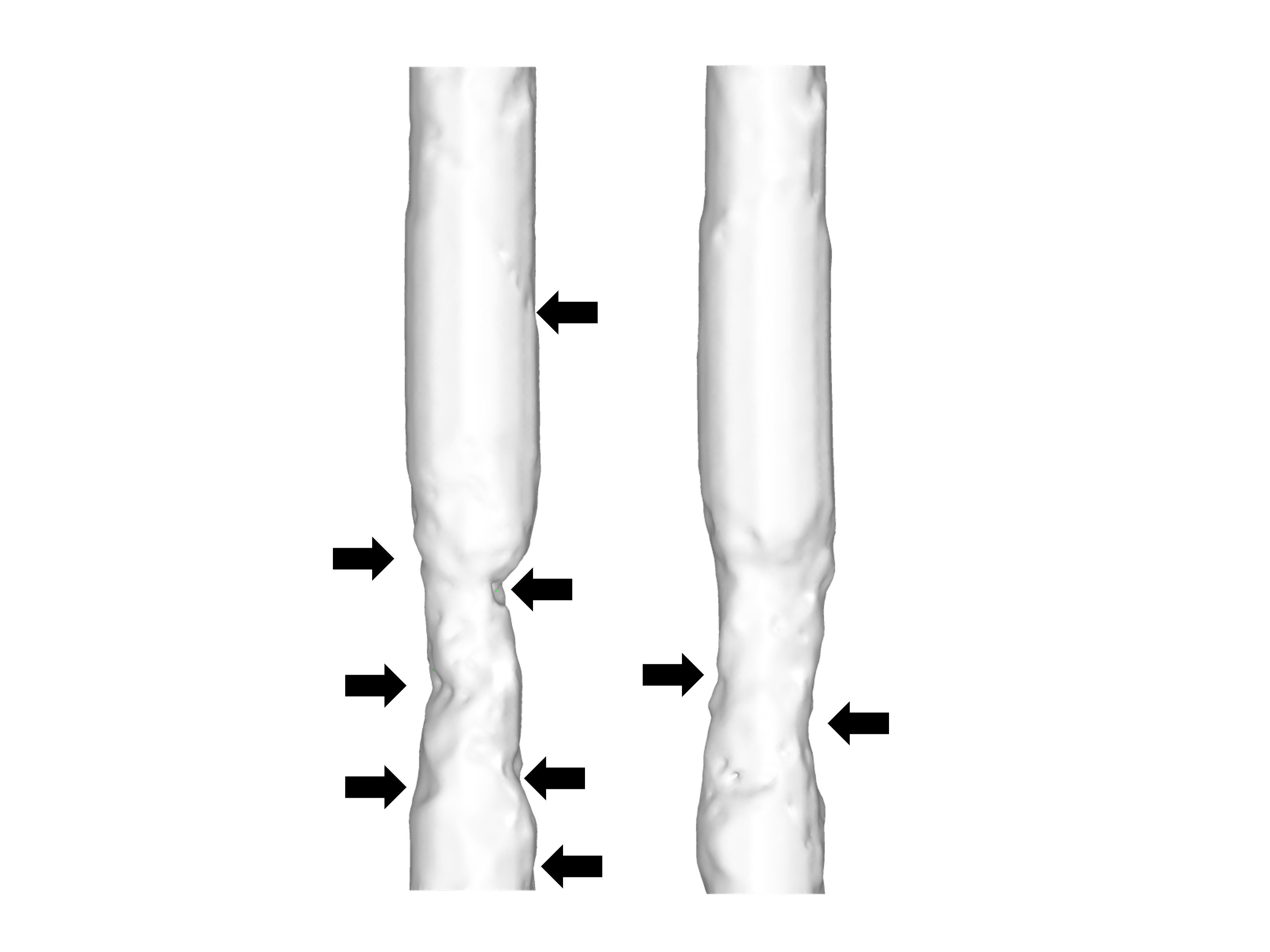}
    \end{subfigure}
    \begin{subfigure}[]{}
        \centering
        \includegraphics[width=0.45\textwidth]{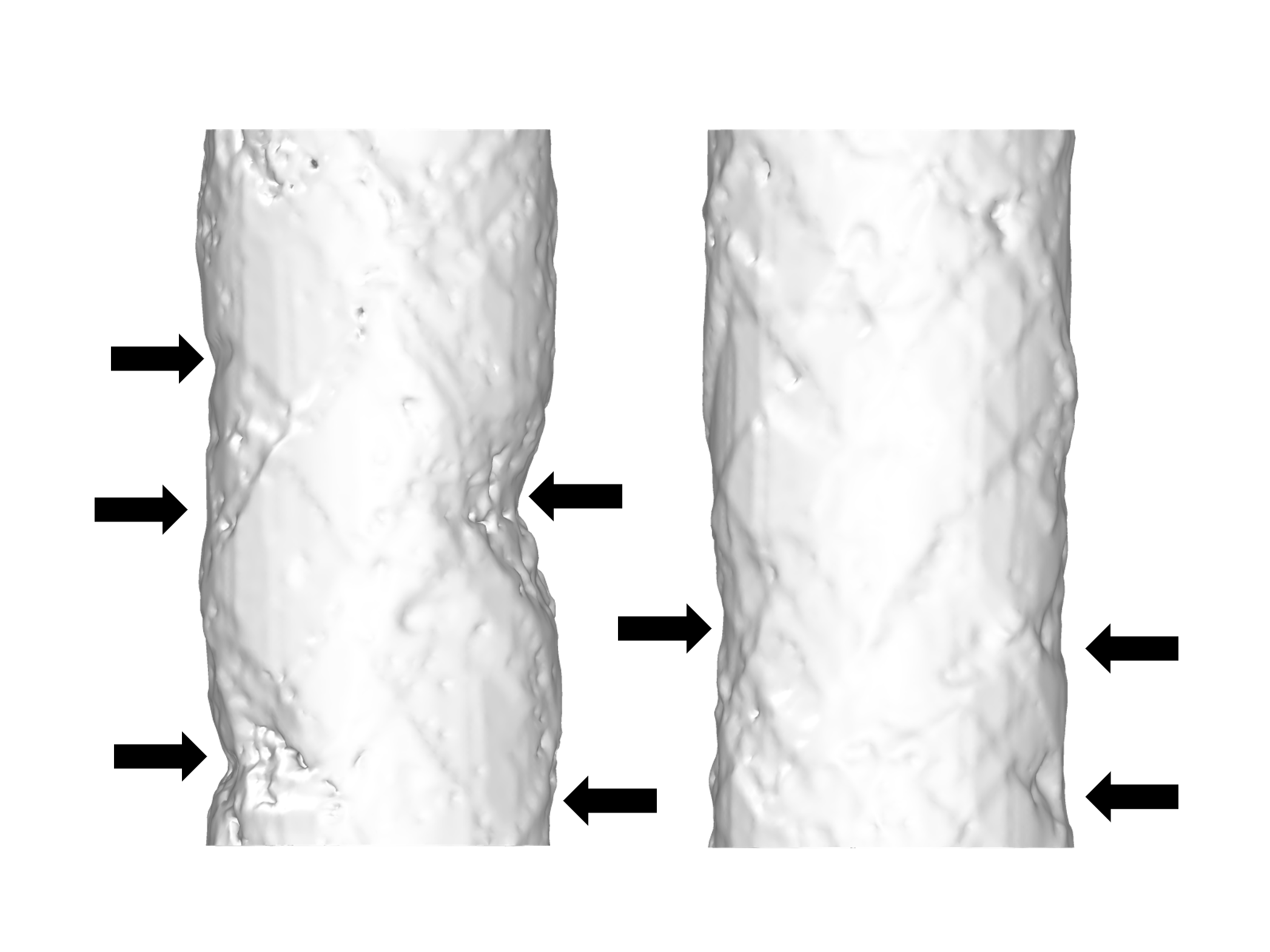}
    \end{subfigure}
    \caption{Stress-stain curves for H-free and H-charged samples with a 6 nm diameter and 30 nm length (a) and a 20 nm diameter and 30 nm length (b). Deformation at 25\% strain for H-free (left) and H-charged (right) samples with a 6 nm diameter and 30 nm length (c) and a 20 nm diameter and 30 nm length (d).}
    \label{FS-SS}
\end{figure}

To investigate the mechanical behavior, the samples were deformed along the Z-axis, and the stress-strain responses of H-free and H-charged samples were compared, as shown in Figure (\ref{FS-SS}a) and (\ref{FS-SS}b). In the H-free samples, the stress values obtained were higher than those of the H-charged samples, indicating greater sample strength. While the stress-strain curve of the H-free samples remains relatively flat after the initial deformation for both sample sizes, the H-charged samples display a sawtooth pattern following the initial deformation. This unique behavior can be attributed to differences in dislocation growth and emission across the samples. Further analysis will focus on understanding the role of dislocation activity in these differences.

\begin{figure}[H]
    \centering
    \begin{subfigure}[]{}
        \centering
        \includegraphics[width=0.45\textwidth]{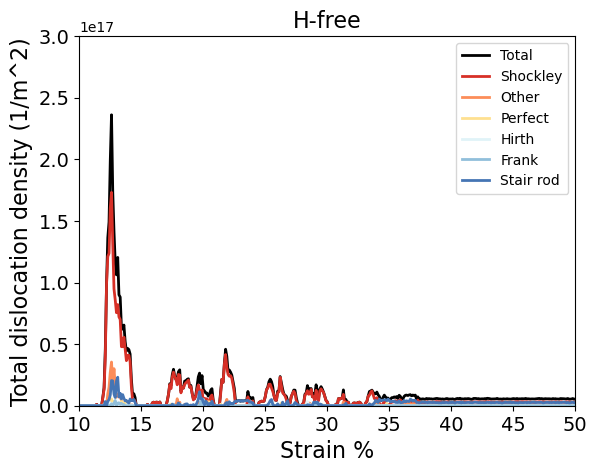}
    \end{subfigure}
    \begin{subfigure}[]{}
        \centering
        \includegraphics[width=0.45\textwidth]{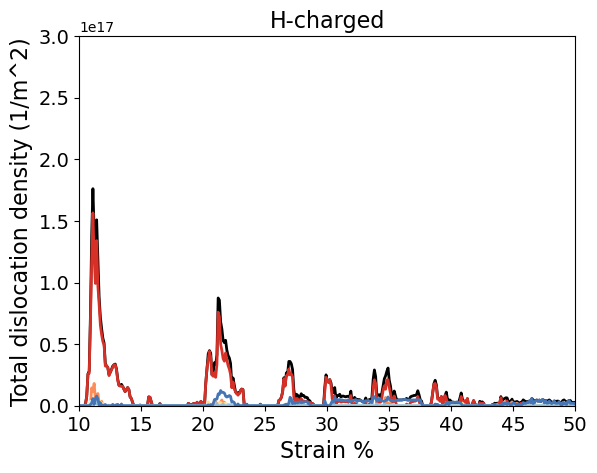}
    \end{subfigure}
    
    \begin{subfigure}[]{}
        \centering
        \includegraphics[width=0.45\textwidth]{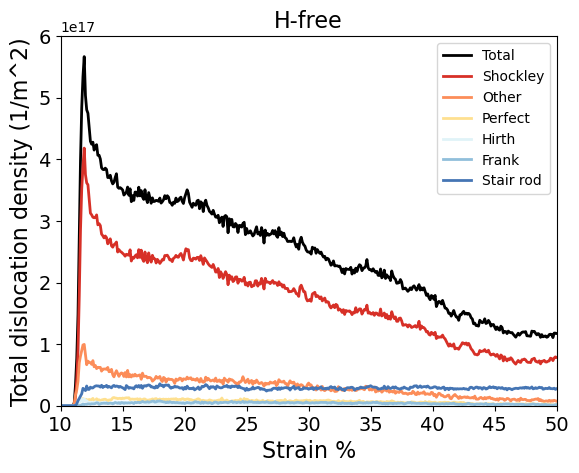}
    \end{subfigure}
    \begin{subfigure}[]{}
        \centering
        \includegraphics[width=0.45\textwidth]{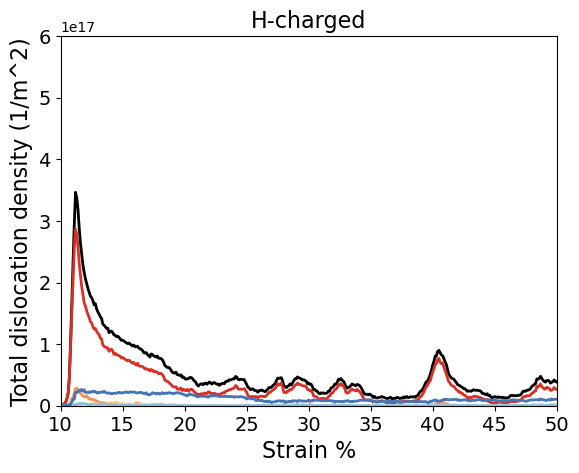}
    \end{subfigure}
    \caption{Distribution of dislocation densities for H-free (a) and (c) and H-charged (b) and (d) samples with a 6 nm diameter and 30 nm length and a 20 nm diameter and 30 nm length respectively.}  
    \label{FS-DXA}
\end{figure}

As depicted in Figure (\ref{FS-DXA}), the H-charged samples exhibit a lower total dislocation density compared to their H-free counterparts. Furthermore, the deformed samples are different in both cases. Figure (\ref{FS-SS}c) and (\ref{FS-SS}d) demonstrates that the H-free samples experience deformation across multiple regions, whereas the H-charged samples show deformation concentrated in a single, localized region. This suggests that H not only influences dislocation density but also affects how deformation is distributed within the material.

\begin{figure}[H]
    \centering
    \begin{subfigure}[]{}
        \centering
        \includegraphics[width=0.45\textwidth]{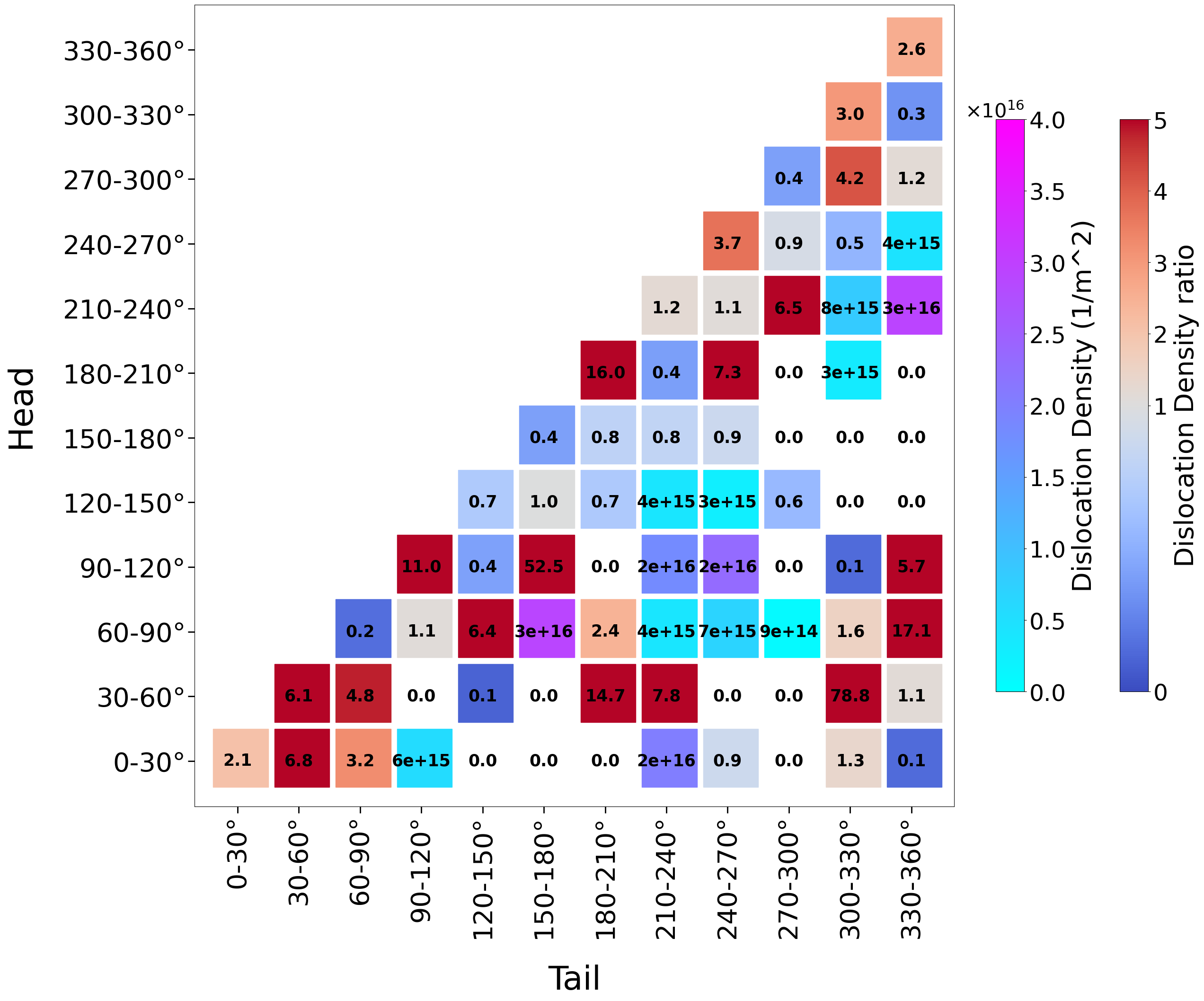}
    \end{subfigure}
    \begin{subfigure}[]{}
        \centering
        \includegraphics[width=0.45\textwidth]{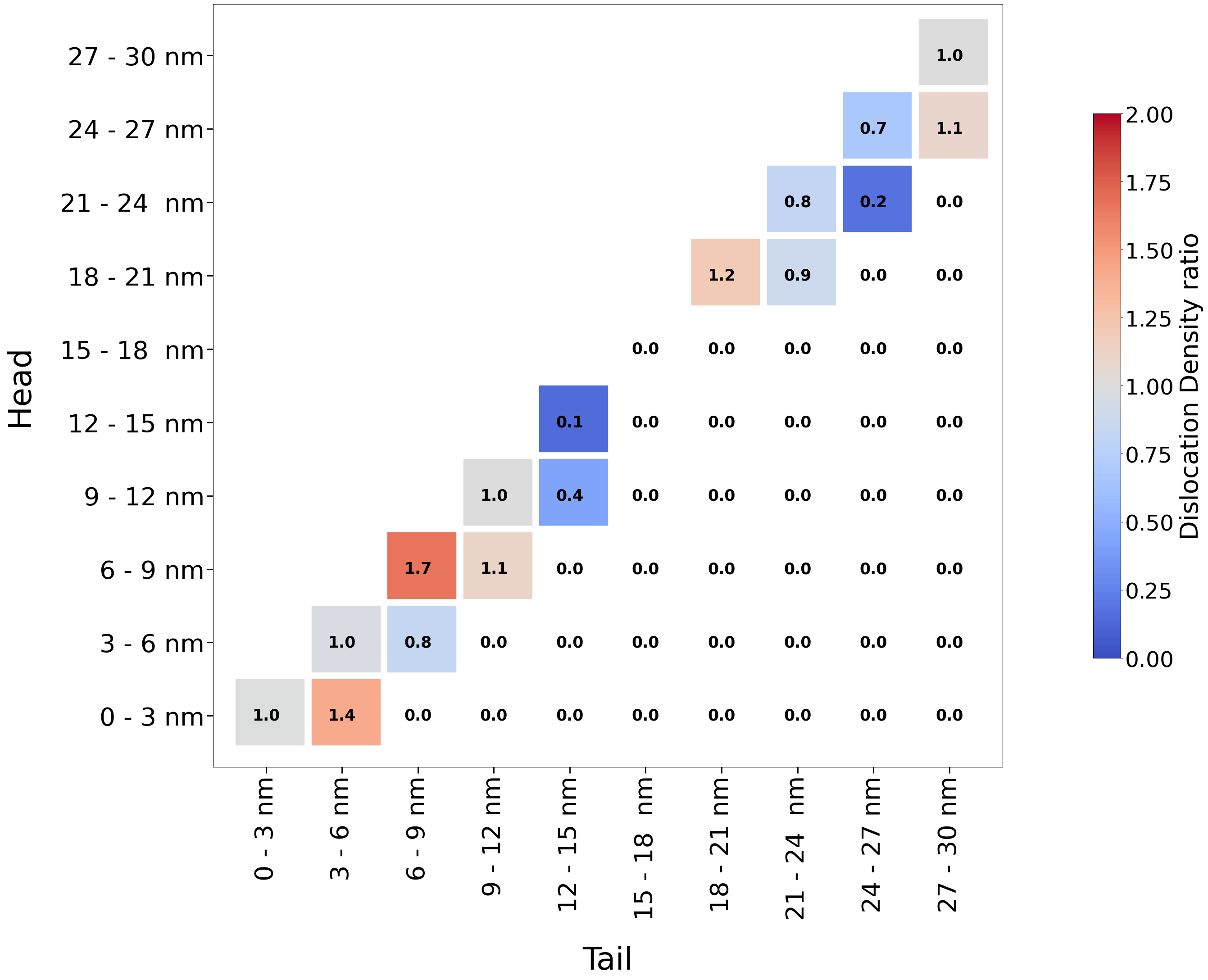}
    \end{subfigure}
    \caption{Normalized dislocation densities as a function of various rotational angles (a) and length regions (b), for the 6 nm diameter and 30 nm length free surface sample.}  
    \label{FS-DXA-emission}
\end{figure}

Dislocations were tracked throughout the deformation process to identify where the head and tail were emitted from the initially undeformed samples. This analysis was conducted in the same subsections as previously defined in Figure (\ref{FS-samples}c). To assess the effect of H on dislocation density in each subsection, the H-charged values were normalized against the H-free values. However, if the dislocation density in the H-free samples was zero, the H-charged dislocation density is presented directly. Figure (\ref{FS-DXA-emission}) shows the values for both the angle and Z-axis. While no clear trend is observed for the rotational angle, the Z-axis plot reveals an increased concentration of dislocation emission in the 6-9 nm regions of the H-charged samples compared to the H-free samples. This region aligns with the localized deformed region shown in Figure (\ref{FS-SS}c) and (\ref{FS-SS}d). The presence of this localized region in the H-charged sample also brings support for the HELP theory.

\subsection{Grain Boundary}
\begin{figure}[H]
    \centering
    \begin{subfigure}[]{}
        \centering
        \includegraphics[width=0.45\textwidth]{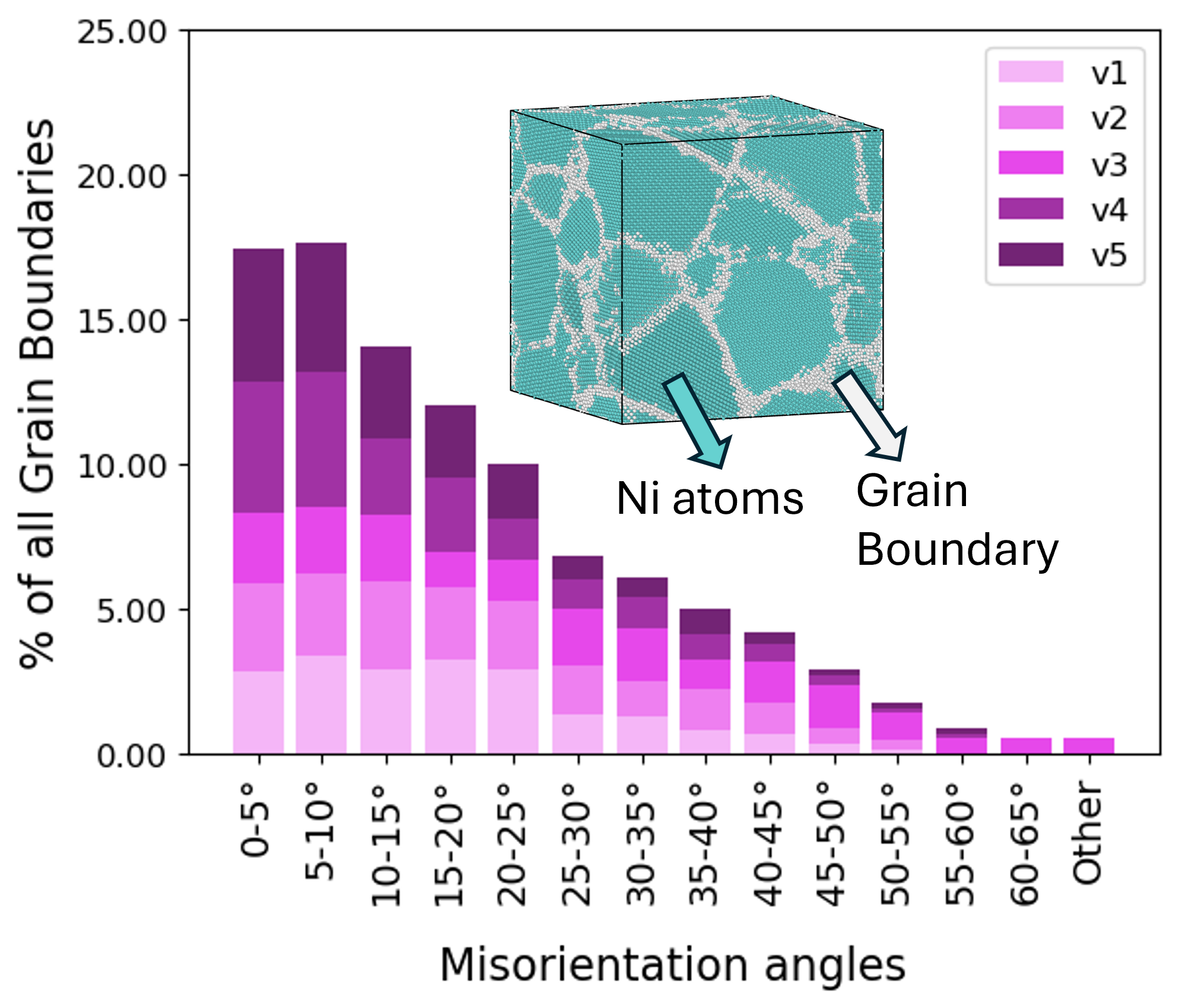}
    \end{subfigure}
    \begin{subfigure}[]{}
        \centering
        \includegraphics[width=0.45\textwidth]{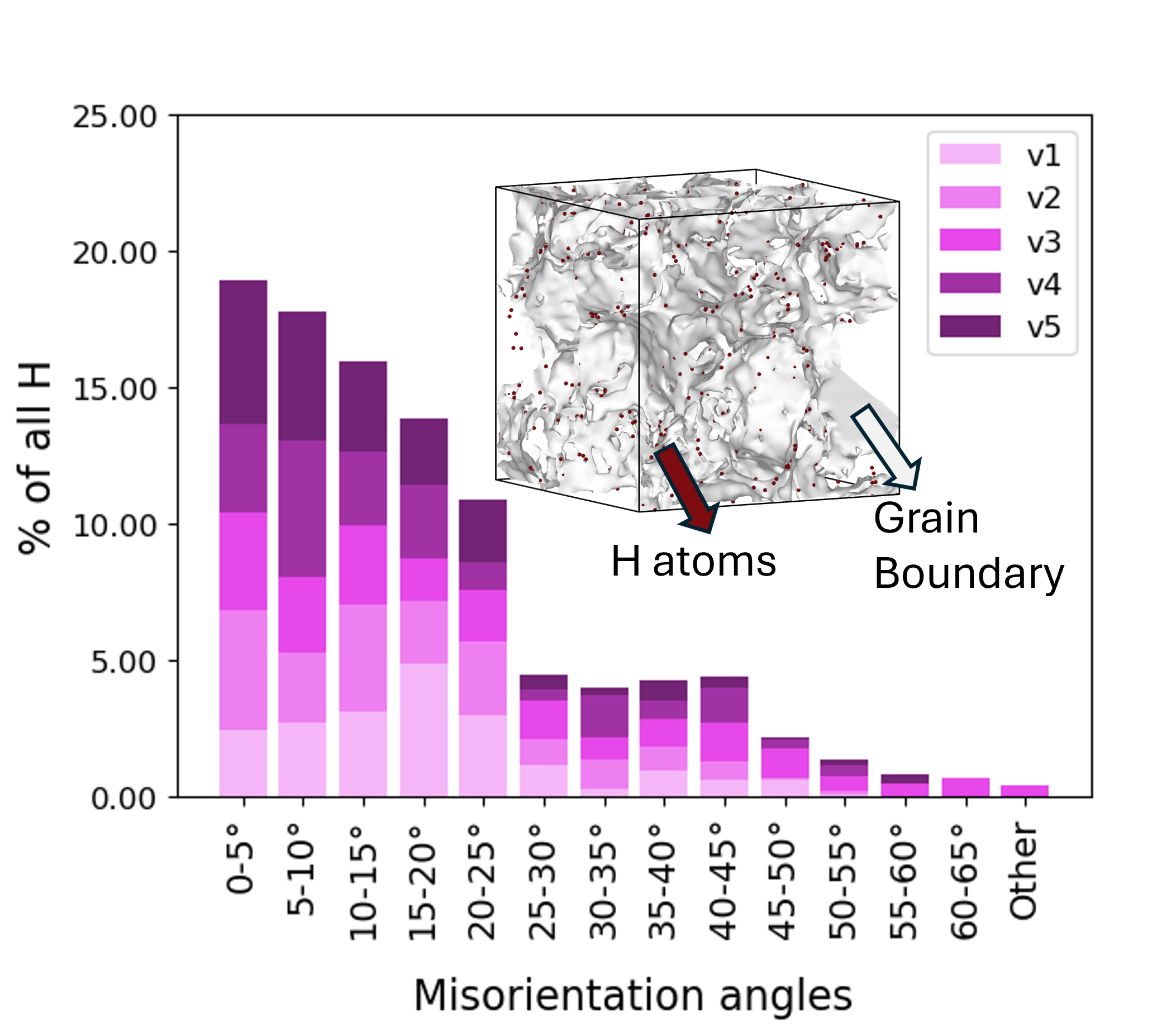}
    \end{subfigure}
    \caption{(a) Distribution of all Grain boundaries according to their misorientation angle. (b) H-segregation tendencies according to grain boundary misorientation angle.}  
    \label{GB}
\end{figure}

As mentioned, five different samples, each containing twenty-seven grains, underwent H charging. In these grain boundary samples, H tends to segregate along the grain boundaries. Specifically, 52.72\% of all H segregates along low misorientation angles. A closer examination of the H-segregation pattern, shown in Figure (\ref{GB}b), reveals three key regions: the first corresponds to misorientation angles $\leq$25$^{\circ}$; the second, misorientation angles 25-45$^{\circ}$; and the third, misorientation angles $>$$45^{\circ}$. Although the transition from low to high misorientation angles occurs at 15$^{\circ}$, providing a well-balanced dataset, a clear preference for H-segregation is observed. Specifically, 77.50\% of all H segregates at misorientation angles $\leq$25$^{\circ}$.

To analyze the mechanical behavior, the samples were deformed along the X-axis, and the average responses of the H-free and H-charged samples were compared. Globally, there is minimal change between the H-free and H-charged curves. However, closer inspection of the plastic region reveals that the inclusion of H results in higher stress within the samples. In these samples, the increased stress in H-charged samples can be attributed to a decrease in dislocation density within the grains. 

\begin{figure}[H]
    \centering
    \begin{subfigure}[]{}
        \centering
        \includegraphics[width=0.45\textwidth]{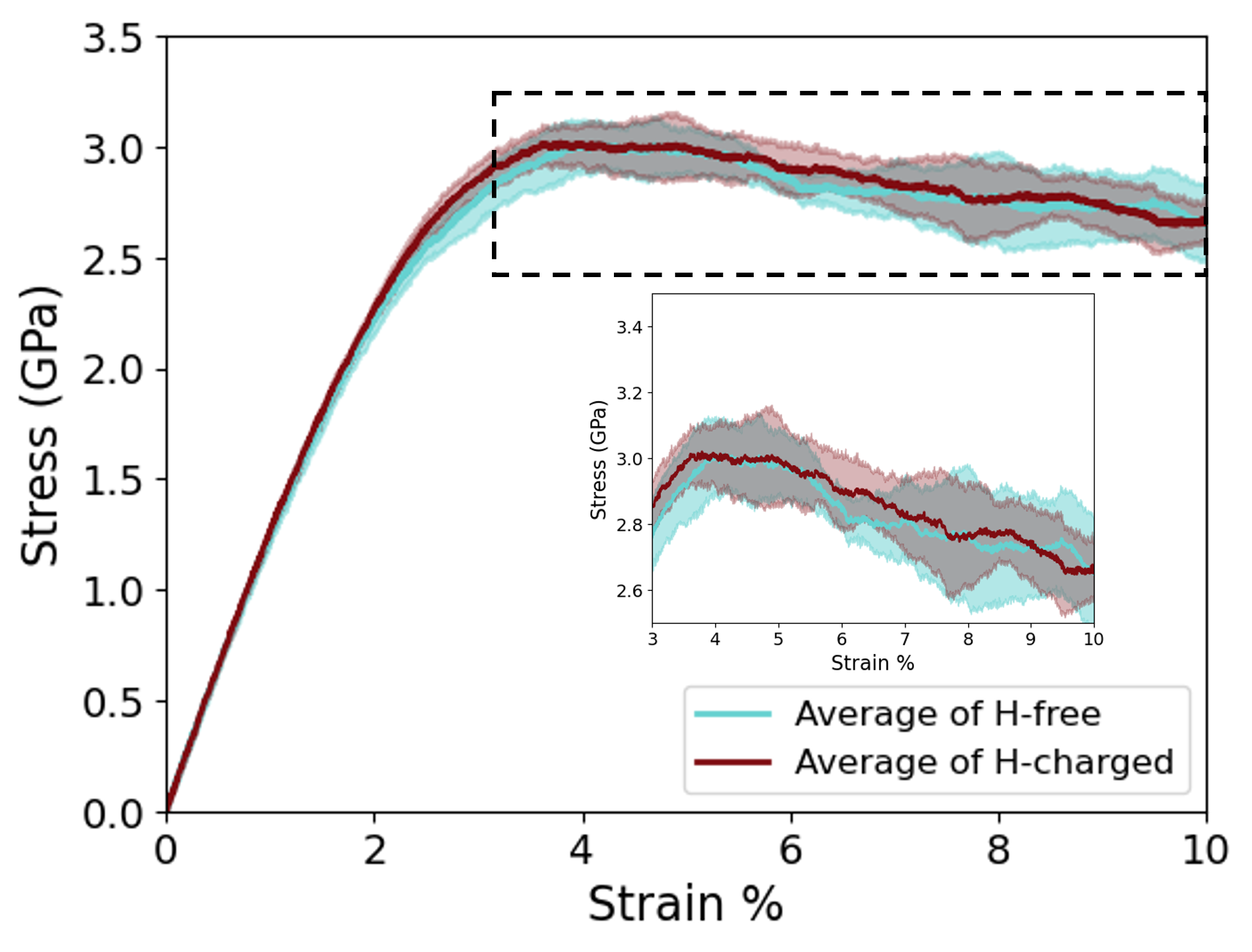}
    \end{subfigure}
    \begin{subfigure}[]{}
        \centering
        \includegraphics[width=0.45\textwidth]{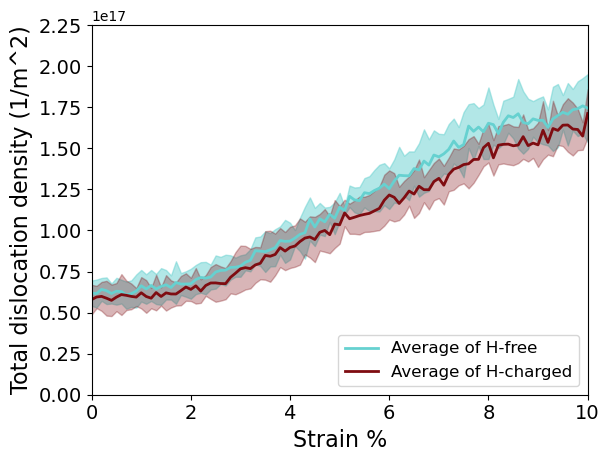}
    \end{subfigure}
    
    \begin{subfigure}[]{}
        \centering
        \includegraphics[width=0.45\textwidth]{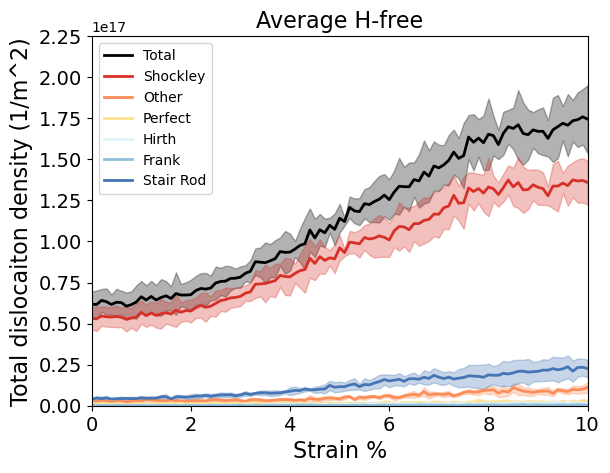}
    \end{subfigure}
    \begin{subfigure}[]{}
        \centering
        \includegraphics[width=0.45\textwidth]{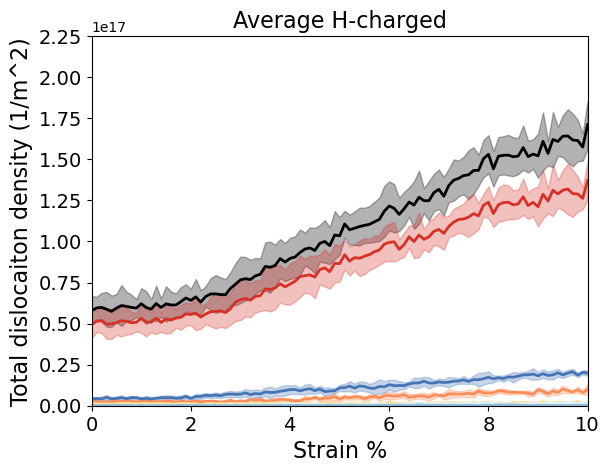}
    \end{subfigure}
    \caption{(a) Average stress-strain curves for each of the grain boundary samples. (b) Average total dislocation density for H-free and H-charged grain boundary samples. Average distribution of dislocation densities for H-free (c) and H-charged (d) grain boundary samples.}
    \label{GB-DXA}
\end{figure}

\begin{figure}[H]
    \centering
    \begin{subfigure}[]{}
        \centering
        \includegraphics[width=0.45\textwidth]{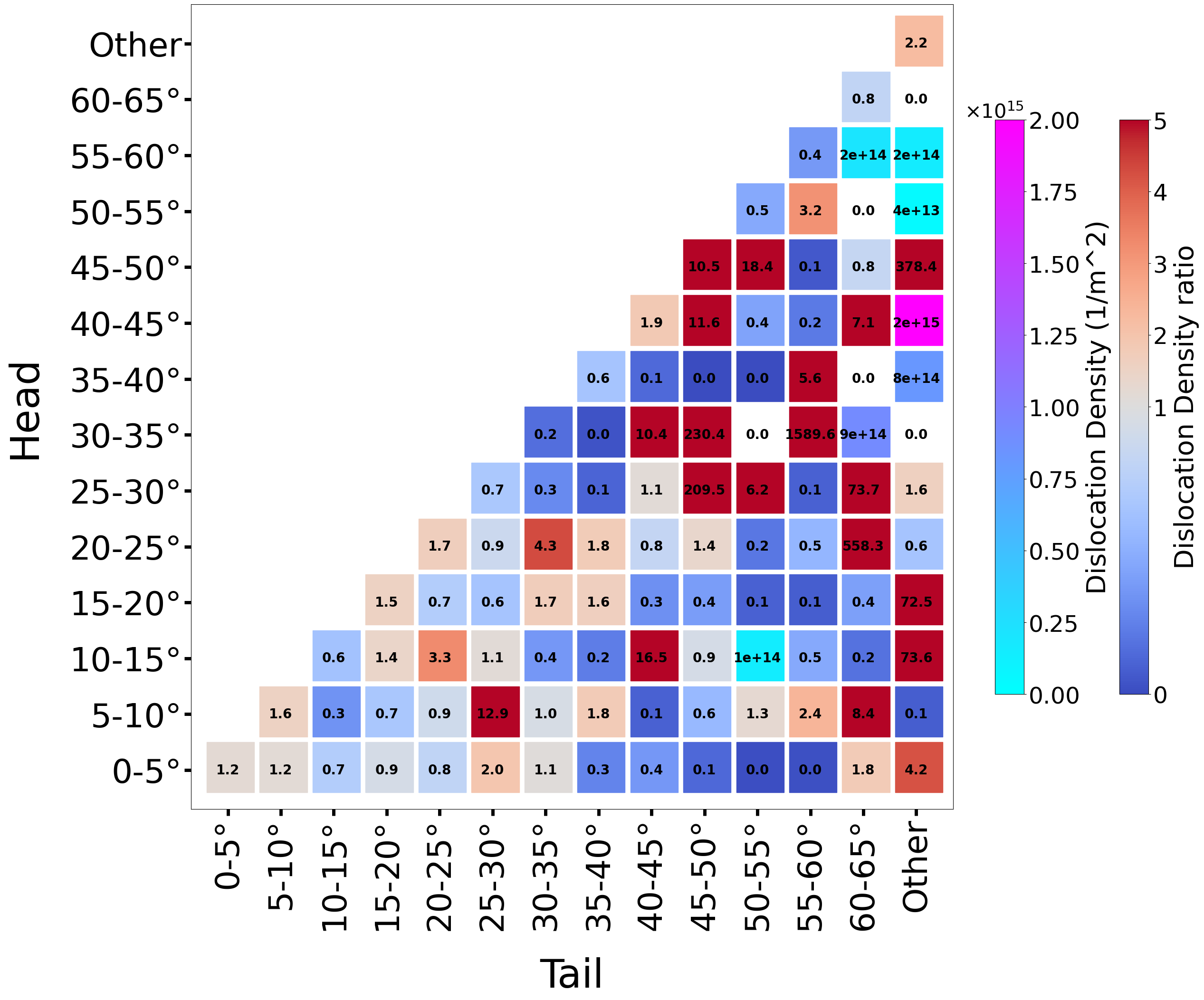}
    \end{subfigure}
    \begin{subfigure}[]{}
        \centering
        \includegraphics[width=0.45\textwidth]{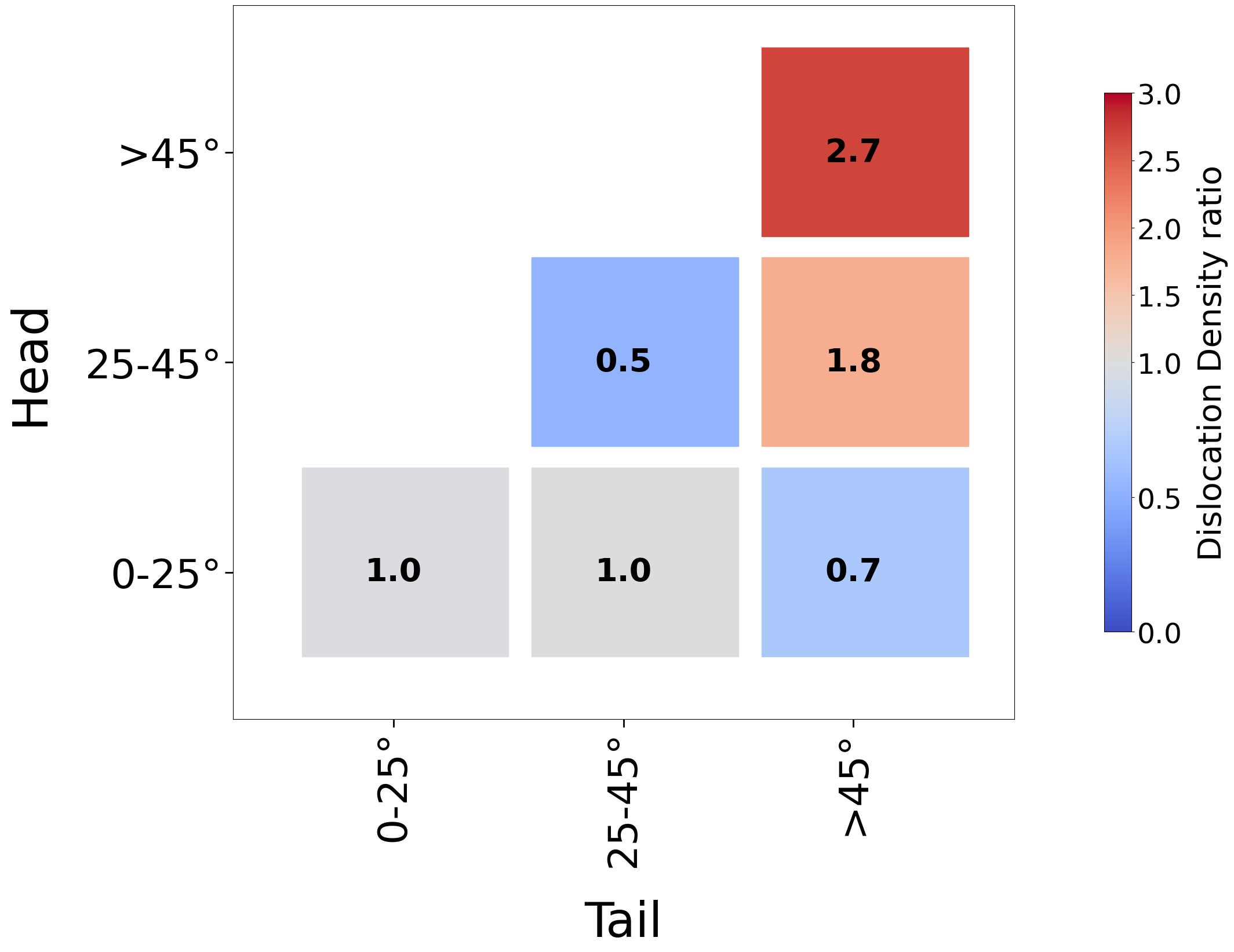}
    \end{subfigure}
    \caption{(a) Normalized dislocation densities as a function of various misorientation angles (b) Simplified version of (a) according to H-segregation regions.}
    \label{GB-DXA-Ratio}
\end{figure}

Similar to how the dislocations were tracked in the free surface samples, the dislocations were tracked to determine which misorientation angle the head and tail were emitted from. Once more, the H-charged values were normalized against the H-free values to assess the effect of H on the average dislocation density. Figure (\ref{GB-DXA-Ratio}) shows these values concerning the misorientation angle. In the misorientation angles $\leq$25$^{\circ}$, there is minimal dislocation density change. However, when looking at the misorientation angles $\geq$25$^{\circ}$ there is a clear increase change in dislocation density. By taking the key regions of H-segregation into account Figure (\ref{GB-DXA-Ratio}a) can be simplified into Figure (\ref{GB-DXA-Ratio}b), by combining the various regions. 

This variation in dislocation density may be due to H altering the activation energy of dislocations. Specifically, H could either increase the activation energy for low misorientation angles to a point where the activation energy for high misorientation angles becomes lower, or it could reduce the activation energy of high misorientation angles to where it is lower than that of low misorientation angles.

\section{Discussion}
In these models, H atoms tend to concentrate in regions where their incorporation into the microstructure requires less energy. In the free surface models, H favors segregation along the surface rather than within the bulk material. At the free surface, atoms are less coordinated with their neighbors compared to the bulk, where atoms are part of a more stable, uniform structure with no gaps in atomic sites. This difference creates a region along the free surface that requires less energy for H to be inserted. Additionally, introducing H increases the local coordination number, which raises the energy required to insert additional H atoms at the same site.

For typical interstitial alloys, atoms are often seen to segregate along high-angle grain boundaries rather than low-angle grain boundaries\textsuperscript{\cite{Kameda1998, Kameda, Chen2, Herbig2013, Kai, Zhou2016, Pavel, Parajuli2019, Tschopp, Schweizer2023, Zhou2023}}. High-angle grain boundaries exhibit greater atomic misorientation and disordered regions\textsuperscript{\cite{Tschopp, Schweizer2023, Zhou2023}}, which create larger interstitial sites and enhanced diffusion paths\textsuperscript{\cite{Schweizer2023,Zhou2023, Pouchon, Zhang2021, Uffelen}}. These characteristics make it easier for interstitial atoms, like Carbon and Nitrogen\textsuperscript{\cite{Kameda1998, Kameda, Chen2, Herbig2013}}, to segregate along high-angle grain boundaries. In contrast, low-angle grain boundaries have a more ordered structure, offering fewer and smaller interstitial sites compared to high-angle grain boundaries. 

However, we hypothesize that H behaves differently from typical interstitial atoms along grain boundaries due to its small atomic size and high diffusivity. Low-angle grain boundaries are composed of ordered structures and stable sites where tiny H can settle at lower energy than in the unstable sites available in disordered regions of high-angle grain boundaries. The sites on high-angle grain boundaries can be more unstable for H, as it can continouously diffuse and relocate to other available sites. 

Experimental validation efforts in the literature have utilized silver-decoration EBSD to investigate H permeation tendencies. In this method, a solution containing silver ions is applied to the surface of the sample. These silver ions are reduced by the H atoms diffusing into the sample, resulting in the formation of silver nanoparticles\textsuperscript{\cite{Nagashima2017, Schober}}. The presence of these silver nanoparticles act as an indirect way of mapping H diffusion pathways into the material. Some studies have shown a greater accumulation of silver nanoparticles along high-angle misorientation grain boundaries compared to low-angle misorientation grain boundaries\textsuperscript{\cite{Moshtaghi2022, Koyama2017, Kim2019}}. In addition, simulations in the literature have explored H diffusion along grain boundaries. These simulations suggest that low-angle grain boundaries impede H diffusion, while high-angle grain boundaries either maintain or accelerate the diffusion\textsuperscript{\cite{Zhou2021, He2021, Luo2021, Kakinuma2024, Xie2020, Zhou2019}}. 

The silver-decoration studies imply that H enters the sample through high-angle grain boundaries, diffusing along these boundaries\textsuperscript{\cite{Moshtaghi2022, Koyama2017, Kim2019, Nagashima2017, Schober}}. However, this experiment does not necessary prove where H finally segregate. When H encounters low-angle grain boundaries, the diffusion of H slows, as corroborated by H diffusion simulations\textsuperscript{\cite{Zhou2021, He2021, Luo2021, Kakinuma2024, Xie2020, Zhou2019}}, and in eventually segregate along these boundaries. Our results thus are not in contradict with these silver decoration experiments as they mainly locate H diffusion paths, and not necessary segregation sites.

Additionally, experimental approaches, such as cryogenic APT, have been used to measure the distribution of H at dislocations, grain boundaries, and precipitates\textsuperscript{\cite{Jiangtao, Barrera2018, Zhou20232}}. This method has successfully identified the presence of H at these defects. However, many of these studies focus on a single high misorientation grain boundary, which may not fully capture the complete H-segregation tendencies when various types of grain boundaries are available\textsuperscript{\cite{Gault2018, Breen2020}}.

Considering dislocation emission, growth, and deformation, the free surface samples analyzed in this study reveal a distinct difference between H-free and H-charged samples. As previously discussed, the H-free samples exhibit multiple regions of deformation, whereas the H-charged samples display a single, localized deformation region. This deformation pattern aligns with experimental observations in H-free and H-charged pentatwinned samples reported by Yin et al.\textsuperscript{\cite{Yin2019}}. In the same study, the presence of H was shown to hinder dislocation nucleation, thereby increasing the energy required for dislocation emission as was calculated by Nudged Elastic Band (NEB) simulations\textsuperscript{\cite{Yin2019}}.

Experimental studies in the literature have employed various techniques, such as environmental transmission electron microscopy (ETEM) and scanning electron microscopy-based electron channeling contrast imaging (SEM-based ECCI), to investigate dislocation activity near grain boundaries under deformation. Both methods, ETEM and SEM-based ECCI, have demonstrated, in the absence of H, dislocations exhibit a certain level of mobility and density\textsuperscript{\cite{Koyama2020, Bond1987, Moshtaghi2022, Bond1988, Bond, Ferreira, Xie20162, Sheit1988, Robertson1986, Xie2021}}. However, with the presence of H, these dislocations displayed increased mobility, suggesting that H plays a promoting role in dislocation mobility and nucleation.

Contrasting to experimental attempts, simulation studies have reported conflicting results regarding dislocation density in the presence of H. Some of the studies indicate a reduction in dislocation density, while others have reported an increase with the presence of H. Some of these discrepancies can be attributed to various factors, including material type, alloying composition, microstructure, H concentration, H diffusivity, applied loading conditions, temperature, stress state, and defect types\textsuperscript{\cite{Koyama2020, Girardin2015, Hachet2019, Yagodzinskyy2010, Girardin2004}}. These contradictions underscore the importance of studying the statistics of H-segregation on fair amount of defects, rather than focusing on a limited window in the microstructure. It is also noteworthy that while the MD simulation results depend on the validity of the potential used, the potential employed in this study has been validated multiple times in previous research\textsuperscript{\cite{Song2013, Koyama2020, Tehranchi2017, Srinivasan2007, Horstemeyer, Srinivasan2005, Chandler2008, Angelo1996, Chandler20082, Drexler2021, Ding2024}}. The obtained results here need to be further validated by DFT calculations. This will be reported in our future studies while we appreciate DFT's limitations in simulating only small and simple case studies of all these defect types. 

\section{Conclusions}
Considering all the results presented in this study and relating them to the HELP mechanism, our simulations offer new insights. In the free surface model, H-free samples exhibit multiple regions of deformation due to higher dislocation emissions. This behavior aligns with observations in H-free samples, where crack blunting occurs as a result of dislocation activity at the crack surfaces, which serve as free surfaces. In contrast, H-charged samples displayed concentrated deformation localized in a single region, forming a singular path for crack propagation. This behavior is consistent with H-charged samples, where the sharpness of the crack surfaces is preserved.

In the grain boundary samples analyzed in this study, observations indicate a reduction in dislocation density. This finding does not support with the notion of increased plasticity in tiny ribbons beneath the crack surface as proposed by the HELP theory. Our simulations show that potentially grain boundaries alone cannot explain the increased plastic deformation. This suggests that the combination of defects, e.g., grain boundaries with free surfaces or vacancies, might be responsible for the enhancement. Our grain boundary study support the decreased plasticity far from crack surfaces in H-charged samples. In conclusion, our studies suggest statistical analyses of H on a more comprehensive defect network to unravel the full picture of the HELP mechanism.


\clearpage
\clearpage

\bibliographystyle{unsrt}
\pagebreak
\bibliography{main.bib}
\end{document}